# Decoupling of Nucleation and Growth of ZnO nano-colloids in solution

Priyanka Sharma, P.B. Barman and Sanjiv Kumar Tiwari[*]

*Department of Physics and Materials Science, Jaypee University of Information Technology*

*Waknaghat, Solan, H.P. 173234, India*

**Abstract:**

In this paper, temporal growth and morphological evolution of ZnO nano-colloids were studied by in-situ UV-Vis absorption spectroscopy and Transmission Electron Microscopy (TEM) respectively. Nucleation of the nanoparticles was observed to occur within 10 sec in the solution after mixing the precursors and there was not any significant change in morphology observed with an increase in growth time. The morphological change was found to depend on interfacial energy curvature. Decoupling of nucleation and growth parameters was observed in the case of the atomically unbalanced reaction while aging of the nanoparticles was found in atomically balanced reaction respectively. The growth of nano-particles was modeled using the Phase-field model (PFM) and compared with the present in-situ growth process.

*Corresponding Author:

E-Mail address: sanjivkumar.tiwari@juit.ac.in

# 1. Introduction

The solution phase route or chemical route has become a very popular and economical method for the preparation of nanoparticles and nano-architectures. Especially, the preparation of ZnO nanoparticles or their nano-architectures is based either on hydrolysis of zinc salt in an organic solvent or the non-hydrolysis method[1,2]. Further, since optical and other physical properties strongly depend on the size of the nanoparticles, the fundamental insight or understanding of the different processes that leads to particle formation is very important. There are three main processes in chemical route synthesis *i.e.* nucleation, growth, and aging [3,4]. Since chemical reactions are acid-base reactions, they are very fast. Hence, compared to nucleation and growth, aging process is relatively slow. Many authors have mainly studied the aging process in detail and there exist very few reports on nucleation and growth. For real-time or in situ monitoring of nucleation and growth, two main experiments have been proposed *i.e.* optical hyper Rayleigh scattering and small-angle X-ray scattering (SAXS) and/or wide-angle X-ray scattering (WAXS). These experiments have a time resolution of few hundreds to thousand milliseconds [5,6]. In hyper Rayleigh scattering, the detection limit of size was 5 nm and SAXS/WAXS were more suitable for metal nanoparticles. Thus, the most effective method for real-time monitoring of nucleation and growth of semiconducting nanoparticles is UV-visible absorption spectroscopy. Lianhua *et.al* has reported the in situ observation of nucleation and growth of CdSe nanoparticles with a fast integration time of the order of 2-4 milliseconds. However, the experiment was carried out using a high-temperature transmission-reflection dip probe immersed in reaction solution [7]. Whereas, in a typical UV-visible spectrograph, the average scan time of any sample ( spectral range of 350-600 nm) is of the order of few minutes providing only the details of growth and aging. Bruno et.al coupled time-resolved X-ray absorption fine structure and UV-visible absorption spectroscopy to monitor the temporal evolution of ZnO quantum dots [8]. They claim that the formation of ZnO colloidal achieves a quasi-steady state equilibrium in less than 3.5 minutes and thereafter, the growth process is controlled by limited aggregation process and kinetics of growth is governed by Ostwald ripening. Therefore, an adequate theoretical model is much needed in order to support the results from UV-visible spectroscopy and clear understanding of growth process and morphological tailoring. Theoretically, nucleation and growth is a well-studied phenomenon. Some well-known methods are classical nucleation theory ( CNT), Ostwald ripening and Lifshitz, Slyozov and Wanger (LSW) theory [2,9–16]. However, each model has its own merits and demerits. For example, CNT fails to predict nucleation rate and dynamical parameters such as kinetic prefactor and interfacial free energy which were used as a fitting parameter. Whereas, LSW model assumes that the transport between the growing particles is diffusion-limited only. To the best of our knowledge, the phase-field crystal (PFC) model is the best known microscopic approach to study nucleation, growth and morphological evolution, because it includes the anisotropies of the interfacial free energy and can be studied in the many order of magnitude of diffusion time scale. PFC method is generalized Phase field theory with crystal order parameter. Thus, just before the initiation of particle growth (where crystal order is absent), the phase field can be well applied to study the growth mechanism.

In the present work, temporal evolution of the growth of ZnO nanoparticles was monitored using in situ UV-visible absorption spectroscopy. Detailed analysis of balanced and unbalanced atomic reactions was performed and the difference between the two was analyzed analytically. Further, phase-field model were employed to model the growth process.

## 2. Material and Methods

Sample preparation involves the simple chemical synthesis where ZnCl$_2$ and NaOH were used as precursors and ethanol (99.9% pure) as a solvent. All the chemicals were used as received. Separate solutions of two different concentrations of ZnCl$_2$ (1mM, 1.25mM) and NaOH (2mM) were prepared in ethanol with a small quantity of water followed by continuous stirring at room temperature. Then both of the solutions (ZnCl$_2$ and NaOH) were mixed slowly to obtain the homogenous colloidal suspension of ZnO nanoparticles [17]. Thereafter two different samples S1 (atomically balanced reaction, 1mM of ZnCl$_2$ and 2mM of NaOH) and S2 (atomically unbalanced reaction, 1.25mM of ZnCl$_2$ and 2mM of NaOH) had been chosen for the study of the growth process. The chemical reactions of the synthesis process are given below:

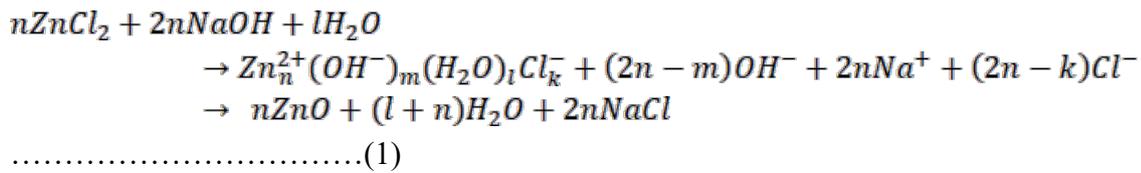

$$nZnCl_2 + 2nNaOH + lH_2O$$
$$\rightarrow Zn_n^{2+}(OH^-)_m(H_2O)_l Cl_k^- + (2n-m)OH^- + 2nNa^+ + (2n-k)Cl^-$$
$$\rightarrow nZnO + (l+n)H_2O + 2nNaCl$$

……………………….(1)

where m,n and k represents the number of moles

To study the temporal evolution of nucleation and growth of sample S1 and S2, the in situ UV-Vis absorption spectra of suspended nanoparticles was recorded at different time intervals (0, 6, 12, 18, 24, 30, 36 min.) using a UV-visible-NIR spectrophotometer ( Perkin Elmer Lambda 750) with a scan time of two minutes in the spectral range of 280-450 nm. Further for morphological evolution Transmission Electron Microscope (TEM: Hitachi -H-7500) images were recorded at the same time interval. The growth of the nano-particles was modeled using Phase-Field modeling (PFM).

## 3. Results and Discussion

In steady-state, allowed direct optical absorption transition between two electronic states is given as $\alpha(h\upsilon) = A(h\upsilon - E_g)^{1/2}$, where $h\upsilon$ and $E_g$ represent the energy of the photon and optical bandgap respectively, $A$ is constant inversely proportional to the index of refraction ($\mu$). However, for transient growth of nanoparticles, $\mu$ is not constant and varies with the progress of time unless the steady-state is reached. In dynamical or in nonequilibrium state, absorption as a function of time is given by the equations :

$$\alpha(t) = \frac{dC_{ZnO}}{dt} = k_{rea} C_{Zncl_2}^{n_1} C_{H_2O/ethanol}^{n_2} \quad \ldots\ldots\ldots\ldots\ldots\ldots(2)$$

$$\frac{d\alpha(t)}{dt} = \frac{d^2 C_{ZnO}}{dt^2} = \frac{dk_{rea}}{dt} C_{Zncl_2}^{n_1} C_{H_2O/ethanol}^{n_2} + k_{rea} \frac{d}{dt}(C_{Zncl_2}^{n1} C_{H_2O/ethanol}^{n_2}) \quad \ldots\ldots\ldots(3)$$

where $C_{ZnO}, C_{Zncl_2}^{n_1}, C_{H_2O/ethanol}^{n_2}$ represents the concentration of ZnO nanoparticles, $ZnCl_2$, $H_2O$/ethanol, $k_{rea}$ is the reaction rate and $n_1$ & $n_2$ are the reaction order respectively.

Therefore, the rate of change of absorption provides more information about the temporal change of reaction rate and reactant concentration both, since $k_{rea}$ changes according to the rate of increase/decrease of temperature given by Arrhenius law [18]

$$\frac{dk_{rea}}{dt} = \frac{E_A}{RT^2} \exp(-\frac{E_A}{RT}) \frac{dT}{dt} \quad \ldots\ldots\ldots\ldots\ldots\ldots(4)$$

where the symbols have their usual meaning.

Figure 1 shows the variation of $d\alpha(t)/dt$ with time for sample S1 and sample S2. Sample S2 was chosen purposely because it has a larger concentration of $Zn^{2+}$ ions as compared to $OH^-$ ions. It is evident from figure 1 that, for sample S1, $d\alpha(t)/dt$ is almost flat/constant which indicates constant aging of the sample, whereas for S2, *dα(t)/dt* first increases up to 20 minutes then decreases drastically. This might be due to excess $Zn^{2+}$ ions present which makes nucleates unstable. Physically speaking, the presence of excess $Zn^{2+}$ ions causes large fluctuations in surface energy. Resultantly, the nucleation reaction gets decoupled from growth after 20 minutes of reaction time. The inset of figure 1, shows typical absorption spectra of ZnO colloidal solution at different times. For colloidal solution, absorption is proportional to the sum of the volume of all the particles which gets a redshift with an increase in time. It is also evident from the figure that half of the absorption maxima ($I_{max}/2$) remain constant with time (marked with dotted arrow) indicating that the particle size distribution remains constant throughout the process.

The average optical energy gap of colloidal ZnO nanoparticles was calculated from the equation $(\alpha h\upsilon)^2 = A\ (h\upsilon - E_g)$. Here a crucial assumption was made that the refractive index of nucleates or particles remains constant with time *i.e.* density of the particles remains constant during the whole process. Particle size was calculated from Brus equation [19,20] given by:

$$E = E_g^{bulk} + \frac{h}{8e}\left(\frac{1}{m_0 m_e} + \frac{1}{m_0 m_h}\right)\frac{1}{r^2} - \frac{1.8e}{4\pi\varepsilon_0 \varepsilon}\frac{1}{r} \quad \text{................(5)}$$

where $E_g^{bulk}$ is the bandgap of bulk ZnO, r is the radius of the colloidal particle, $m_e$ and $m_h$ are an effective mass of the electron (0.26) and holes (0.59), ε is the relative permittivity of ZnO (8.5), $\varepsilon_0$ is the permittivity of free space (8.85×10$^{-12}$ F/m) and $m_0$ = 9.1×10$^{-31}$ kg.

Taking the derivative of Eqn. (5) *w.r.t* time;

$$\frac{dE}{dt} = -\frac{h\pi}{3e}\left(\frac{1}{m_0 m_e} + \frac{1}{m_0 m_h}\right)\left(\frac{1}{\frac{4}{3}\pi r^3}\frac{dr}{dt}\right) + \frac{1.8e}{\varepsilon_0 \varepsilon}\left(\frac{1}{4\pi r^2}\frac{dr}{dt}\right) \quad \text{…………………….(6)}$$

The above equation implies that for zero slopes ($dE/dt = 0$) or in steady-state, the surface to volume ratio of particles remains constant. However, for finite $dE/dt$ values the first term indicates volumetric dependence and the second term indicates surface dependence. Figure 2, shows the variation of $dE/dt$ with time for sample S1. For $dE/dt = 0$ indicates constant surface to volume ratio or the aging of the sample, which is in accordance with the absorption data (figure 1). For sample S2, $dE/dt$ is initially positive till 20 minutes and then after turns negative, which implies surface-mediated growth and gradual decoupling of nucleation and growth parameters respectively. The inset of figure 2 shows the variation of bandgap with time, which is in accordance with Brus equation (5). Figure 3, shows the variation of particle size with time for S1 and S2 (inset). Experimental data points were fitted to the equation:

$$d = d_0 + kt^p \quad\quad\quad\quad\quad\quad\quad\quad\quad (7)$$

where $d, d_0$ and $k$ represent the particle size, initial particle size and proportionality constant respectively. The parameter $p = 1/3$ indicates purely diffusion-controlled growth for uniformly distributed colloidal particles.

Blue and green dotted lines in figure 3 represent the function plot of $d = d_0 + kt^n$, for $p = 1/3$ and $1/2$ respectively. For sample S1, fitting parameters are $d_0 = 1.61 \pm 0.06$ nm, $p = 0.37 \pm 0.06$ and $k = 0.11 \pm 0.03$ with fitting accuracy of 98%, whereas for sample S2, parameters are $d_0 = 1.78 \pm 0.05$ nm, $p = 9.99 \pm 1.72$ and $k = 6.28 \times 10^{-16}$ respectively. This function plot for sample S2 reveals completely different behavior in terms of the nature of the radius of curvature. Here, the proportionality constant $k$ depends on interfacial energy. Diffusion of nucleates present at solid-liquid interface depends on the local curvature of interfacial energy which induces a difference in local equilibrium concentration and hence provides a driving force for the growth of larger particles [21]. In S2, the local thermodynamical fluctuations induce large perturbation in local interfacial energy curvature which decouples nucleation rate and growth rate kinetics. Because of thermodynamical fluctuations and excess $Zn^{2+}$ ions induce interfacial energy curvature, sample S2 shows the unexpected value of k. To visualize the effect of interfacial energy curvature on the morphology of colloidal particles of S2, nanoparticles were deposited on the TEM grid after sonication. Figure 4 shows the TEM image for different times. Although, it seems that a large number of particles are clustered together, but clear variation in morphology can be noticed in figure 4(e) and 4(f). The inset of figure 4 shows log-normal particles size distribution. Log-normal distribution becomes very narrow in figure 4(f), which is a visual signature of decoupling of nucleation rate and growth rate kinetics. Figure 5 shows the cumulative log-normal distribution of particle size and the inset shows the variation of average particle size (calculated from TEM image) with time. The inset figure indicates a decrease in particle size after 30 min due to decoupling of nucleation and growth. However, the size calculated from TEM is larger than the size calculated from the Brus equation because of clustering.

## 3.1. Modelling of growth

The chemical reaction for the synthesis of the ZnO nano-particle in the solution phase is represented in equation 1. In equation 1, the zinc-ligands molecules act as a precursor for nucleation, and for 2n = m+k represents zero charge precursor and homogeneous nucleation. In the present case, two concentrations of zinc salt were chosen, one for S1 and the other for S2 respectively. In the case of homogeneous nucleation, due to the concentration gradient of precursors, NaOH will diffuse into the $ZnCl_2$ solution. Qualitatively, the first diffusion occurs and then chemical reaction starts. For complete reaction to occur, ions must diffuse at least to the order of the radius of the added droplet, hence $r_{droplet} = (6Dt)^{1/2}$ (where $r_{droplet}$ = radius of the droplet, D = diffusion constant and t = time). Typically 2ml of NaOH droplet has a radius of the order of 0.7 mm ($4/3 \pi r^3_{droplet}$ = 2 ml) since, D ~ 1.3 x $10^{-5}$ $cm^2$- $sec^{-1}$, hence t ~ 10 sec.

This simple calculation reveals that ions will take about 10 sec to fully diffuse into the $ZnCl_2$ solution. However, at the same time $OH^-$ ions present at the surface will react with Zn ions and subsequently form $Zn(OH)_2$ which acts as a nucleation center for further growth. Therefore, it is quite logical to assume that the first nucleation center forms and growth starts thereafter.

From the above calculation, it is evident that the diffusion of NaOH into $ZnCl_2$ is slow compared to their chemical reaction at the surface. Therefore, it is assumed that a local equilibrium is created between free NaOH (in bulk) and immobilized component $Zn(OH)_2$ at the surface of the drops. In this case, the concentration of $Zn(OH)_2$ or nucleates must be correlated to the concentration of NaOH.

We assume $C_{Zn(OH)_2}$ = R $C^n$ $_{NaOH}$, where R = constant and n = 1,2,3......

Physically, R represents nucleation rate in chemical diffusion reactions. If R is very large then nucleation rate or formation of $Zn(OH)_2$ is also very large as compared to temporal decay of initial NaOH. For n = 1, S = R C (where $C_{Zn(OH)_2}$ = S), the diffusion equation is given by:

$$\frac{dC}{dt} = D\frac{d^2C}{dx^2} - \frac{dS}{dt}, \text{ for linear equilibrium} \ldots\ldots\ldots\ldots(8)$$

$$R\frac{dC}{dt} = D\frac{d^2C}{dx^2} - \frac{dC}{dt} \text{ , or} \ldots\ldots\ldots\ldots\ldots\ldots\ldots\ldots\ldots(9)$$

$$\frac{dC}{dt} = \frac{D}{R+1}\frac{d^2C}{dx^2}, \ldots\ldots\ldots\ldots\ldots\ldots\ldots\ldots\ldots\ldots\ldots\ldots\ldots(10)$$

This means that in the overall process of diffusion, the diffusion constant(D) will reduce by a factor of R+1 if the reaction involves. Similarly, for n = 2, the diffusion equation(10) with reaction will be modified to;

$$\frac{dC}{dt} = \frac{D}{2R+1}\frac{d^2C}{dx^2},  \quad\quad\quad\quad\quad\quad\quad\quad\quad\quad\quad\quad (11)$$

and D will be reduced by a factor of 2R+1. This indicates that the ideal situation is to assume the existence of a large number of nucleation centers for the further growth process. Therefore, the formation of ZnO nano-colloidal in suspension is due to non-separable events of diffusion and reaction both with effective diffusion constant $D(R+1)^{-1}$ and $D(2R+1)^{-1}$ for n = 1 and 2 respectively. It is important to note that, in this particular diffusion-reaction process, we did not consider local reaction or any nonlinear combination of $C_{NaOH}$. This is because, local reaction and diffusion-reaction are of the form $\frac{dC}{dt} = D\frac{d^2C}{dx^2} + R(C)$, where R(C) accounts for all local reactions used in the field of pattern formation given by the Fisher-Kolmogorov-Petrovsty-Piskunov equation[22]. Moreover, a different form of R(C) yields different equations. For example for R(C) = C(1-C), R(C) = C(1-$C^2$) and R(C) = C(1-C)(C-α) with 0<α<1 corresponds to Fisher's equation, Newell-Whitehead Segel equation, and Zeldovich equation respectively [23]. These equations were used to describe the spreading biological population, Rayleigh Benard convection and combustion theory respectively.

Therefore, further, we assume that all nucleation centers have been formed and are available for the further growth process. Physically, a solution with a nucleation center can be thought of or visualize as ordered phase (nucleation centers), disordered phase (liquid) and interface of nucleates and liquid or interface of order and disorder phase respectively. A significant spatial variation of phase is assumed to be larger than the width of the interfacial region. This is a very important assumption because the constant interface will eventually evolve into spherical particles whereas; spatial variation of the interface will give rise to different morphology which is the main reason behind the different reported morphology such as nano-flower, nano-flakes, *etc*. [24].

Since all the nuclei had already been formed, only growth is taken into account (there will not be any involvement of latent heat). Moreover, growth is also responsible for continuous phase transformation. Temporal evolution of solid phase is given by Cahn-Hilliard (CH) equation[25,26]

$$\frac{\partial \phi}{\partial t} = \nabla.(M\nabla \frac{\delta F}{\delta \phi}), \text{ and } \mu = \frac{\delta F}{\delta \phi} \text{ or } \frac{\partial \phi}{\partial t} = \nabla.M\nabla \mu \quad \text{...............(12)}$$

where ϕ, M, μ, and F represent phase, mobility of solute, chemical potential and total free energy functional respectively. Here order parameter ($\phi$) is a conserved quantity. Total free energy functional is given as

$$F(\phi,T) = \int_V \left\{ \frac{1}{2} |W_0 \nabla \phi|^2 + f(\phi(x),T(x)) \right\} d^3x \quad \text{...............(13)}$$

where $f(\phi,x) = \frac{\varepsilon}{2a^3} \phi(1-\phi) + \frac{k_B T}{a^3} \phi(x) \ln \phi + (1-\phi)\ln(1-\phi)$ …………………..(14)

Here, '*a*' is the distance between two discrete points and we assume temperature is constant and no nearest-neighbor interaction. $W_0$ is a coefficient that includes surface energy because it multiplies a gradient of order parameter which varies only at the interface. For simulations we choose free energy function f(ϕ) given as:

f(ϕ) = Aϕ²(1-ϕ)² ……………………(15)

where A is free energy constant.

Using equation (13), equation (12) governing for the temporal evolution of phase can be modified as

$$\frac{\partial \phi}{\partial t} = \nabla . M \nabla (h - 2\kappa \nabla^2 \phi) \text{ or } \frac{\partial \phi}{\partial t} = M \nabla^2 (h - 2\kappa \nabla^2 \phi) \quad \text{...........(16)}$$

Here, $h = \frac{\partial f(\phi)}{\partial \phi} = 2A\phi(\phi-1)(2\phi-1)$, $\kappa$ is the gradient energy coefficient which represents interfacial energy. Fourier transformation method is being used to simulate the transformation equation (16) to equation (17) given below.

$$\frac{\partial \{\phi\}_k}{\partial t} = -Mk^2 (\{h\}_k + 2\kappa k^2 \{\phi\}_k) \quad \text{...........(17)}$$

where $\{.\}_k$ indicates the spatial Fourier transform of the quantity $\{.\}$ and $k$ is the Fourier vector. The semi-implicit discretization of equation (17) is given by:

$$\frac{\phi(k,t+\Delta t) - \phi(k,t)}{\Delta t} = -Mk^2 \{h\}_k - 2\kappa Mk^4 \phi(k,t+\Delta t)$$
$$\phi(k,t+\Delta t) = \frac{\phi(k,t) - Mk^2 \{h\}_k \Delta t}{1 + 2\Delta t \kappa M k^4} \quad \text{...........(18)}$$

Here, $\Delta t$ represents the time step for numerical integration. For simulation we have used scaled parameters, grid size scaled to interfacial width W, time step to $\tau$, free energy functional to $k_B T$. Scaled parameters make free energy constant (A) and gradient energy dimensionless. So, we have mainly four scaled parameters used for computation *i.e.* initial concentration of nucleates (C), free energy constant (A), mobility (M) and interfacial energy ($\kappa$). However, a very small noise term (0.002) was added to C to introduce spatial fluctuation in C. Two different set of parameters *i.e* ( C = 0.25, A = 2, M = 0.5, $\kappa$ = 0.5 and C = 0.30, A = 1, M = 1, $\kappa$ = 0.5) were used and size evolution were monitored at different time step $\Delta t$. Figure 6 and figure 7 shows simulated image for two different set of parameters at time setps

of Δt=100,500,1000,1500 and Δt=100,900,2100,3000 respectively. It clearly reveals that nucleates and interfacial curavature are well formed and visisble by increasing the inital concentration at Δt=100 steps. Inset of figure 6 and 7 indicates log normal distrution of averarage particle size.

## 4. Conclusions

In the present study, ZnO nano-colloids were prepared by using $ZnCl_2$ and NaOH with two different concentrations. The ZnO nano-colloids were continuously studied for nucleation and growth at different time intervals. The nucleation and growth process of the ZnO nano-colloids were investigated using the PFM model. The PFM model successfully explains the evolution of nanoparticles with the increase in time steps. It was concluded experimentally that particles nucleate within 10s in the colloidal solution followed by their growth then after. Further, the Cahn-Hilliard (CH) equation explains the temporal growth of the particles with the assumption of a constant continuum term ($\phi$). However, in case of atomically unbalanced reaction, the traces of decoupling of nucleation and growth were found in UV-Vis absorbance spectra. Additionally, the simulation studies showed that the spherical particles were due to constant interfacial term ($\kappa$). Morever, there was a redshift in absorbance maximum with respect to growth time was observed.

Thus in the present paper, temporal growth of ZnO nano-colloids were studied. The findings from the current study were believed to be useful in materials science and device fabrication process. The interfacial energy term ($\kappa = 0.5$) was considered to be constant in the present case. However, the variations in $\kappa$ were believed to be responsible for various microstructure evolution *viz*. nanowires, nano-flowers, nano-flakes and nano-helical, *etc*.

## 5. Acknowledgement

The authors would like to thanks Dr. Pankaj Sharma (Applied Science Department, National Institute of Technical Teachers Training & Research, Chandigarh) for sample characterizations and guidance.

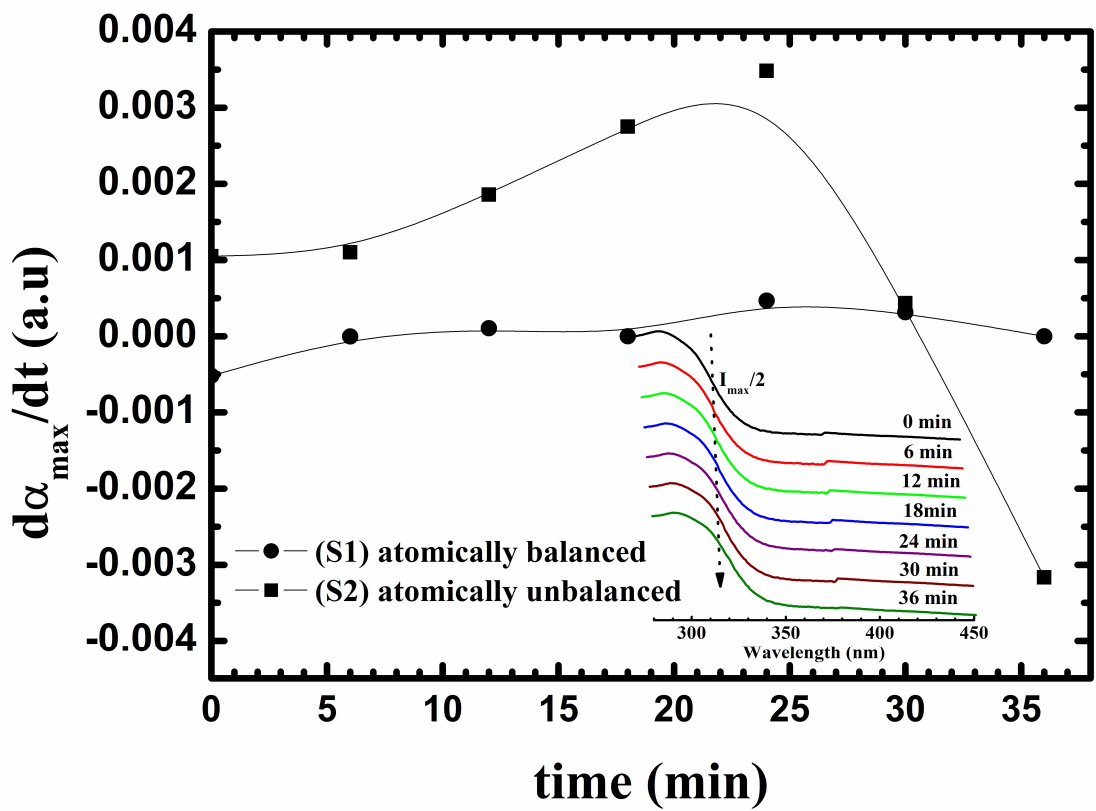

Figure 1. Variation of absorption rate with time, inset shows typical absorbance spectra of the ZnO colloidal particles in the solution at different growth times for the sample S1.

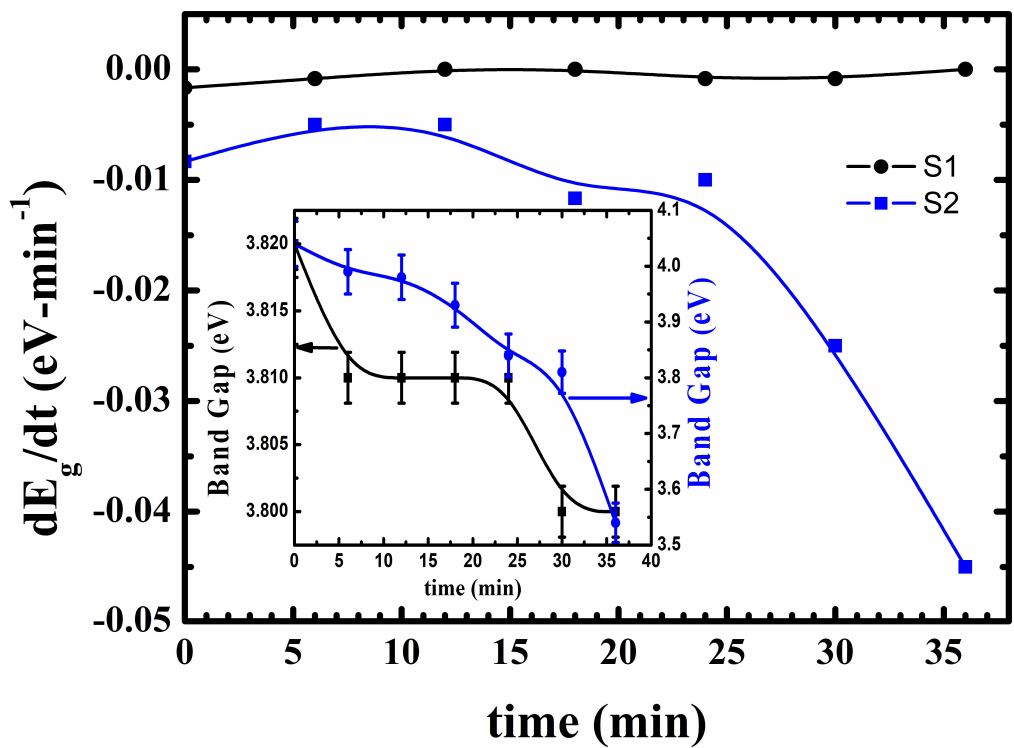

Figure 2. Variation of the rate of bandgap with time, inset shows the variation of bandgap calculated from Tauc plot with time.

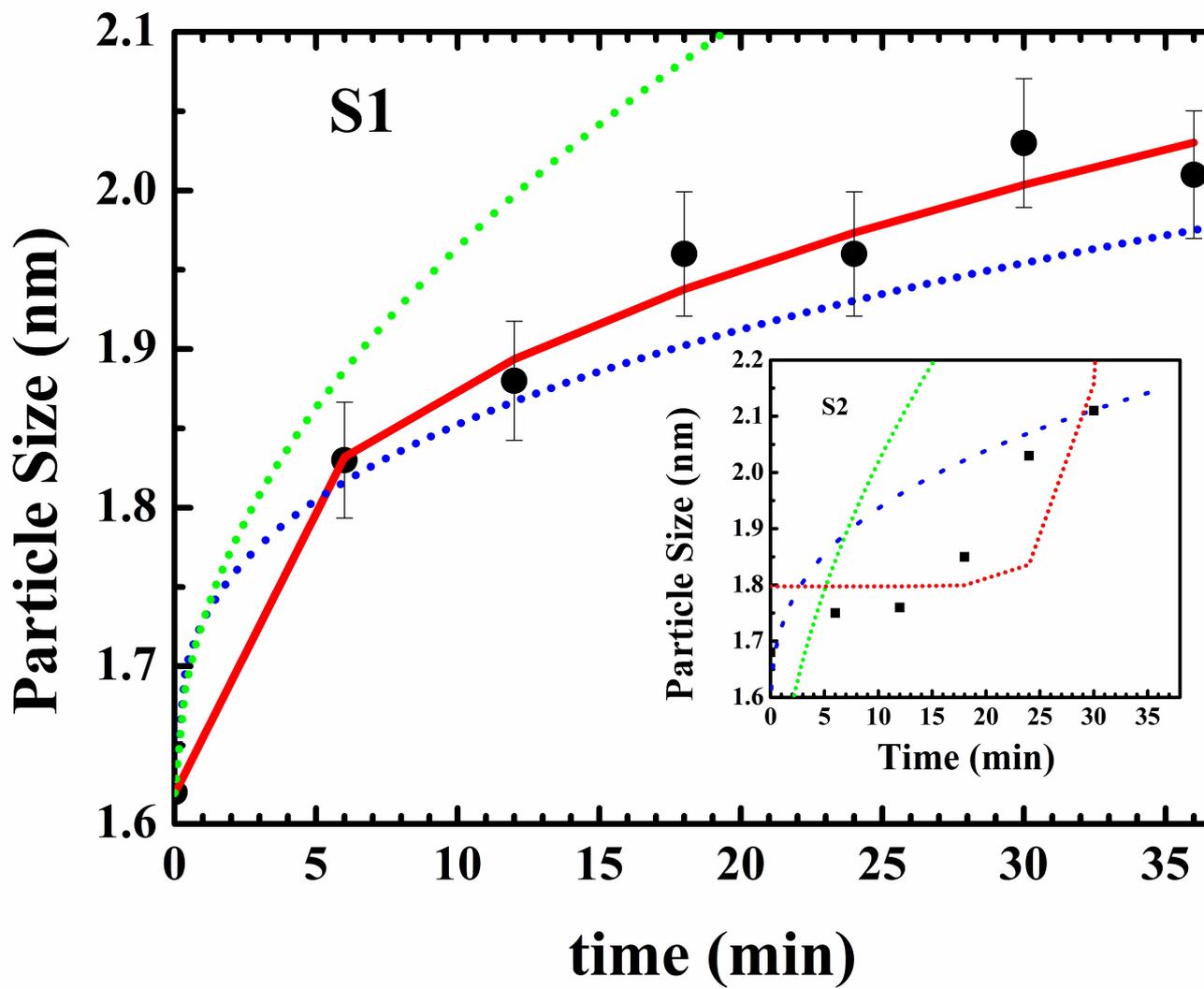

Figure 3. Variation of particle size with time for sample S1, solid red line represents theoretical fit ( $d = d_0 + kt^p$ ) to experimental data points with p = 0.37, blue and green dotted

lines shows function fit with p = 1/3 and 1/2 respectively, inset shows the variation of particle size with time for sample S2.

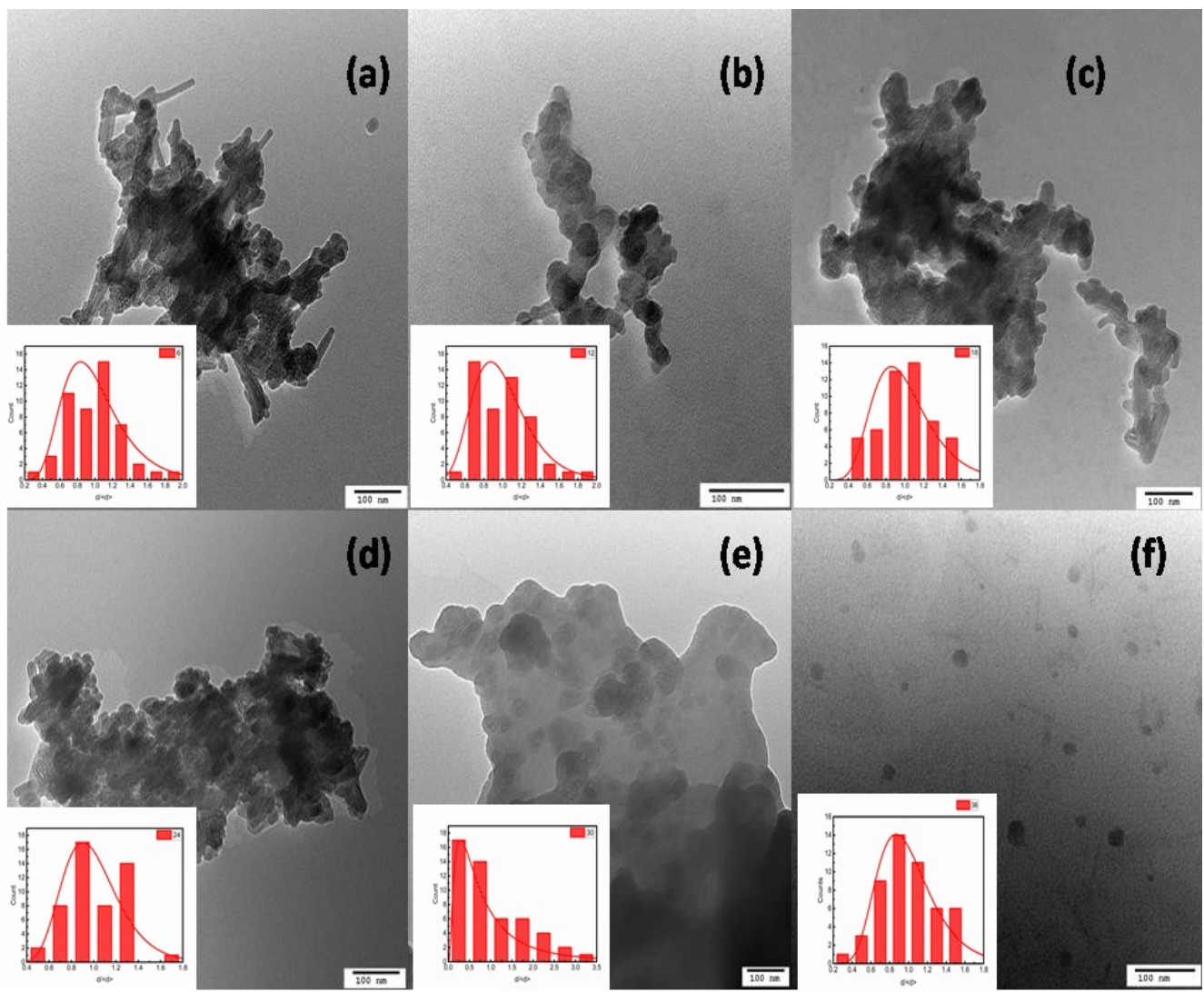

Figure.4 TEM images with the log-normal distribution (in inset) of the ZnO nano-colloids for different times obtained after sonication for sample S2.

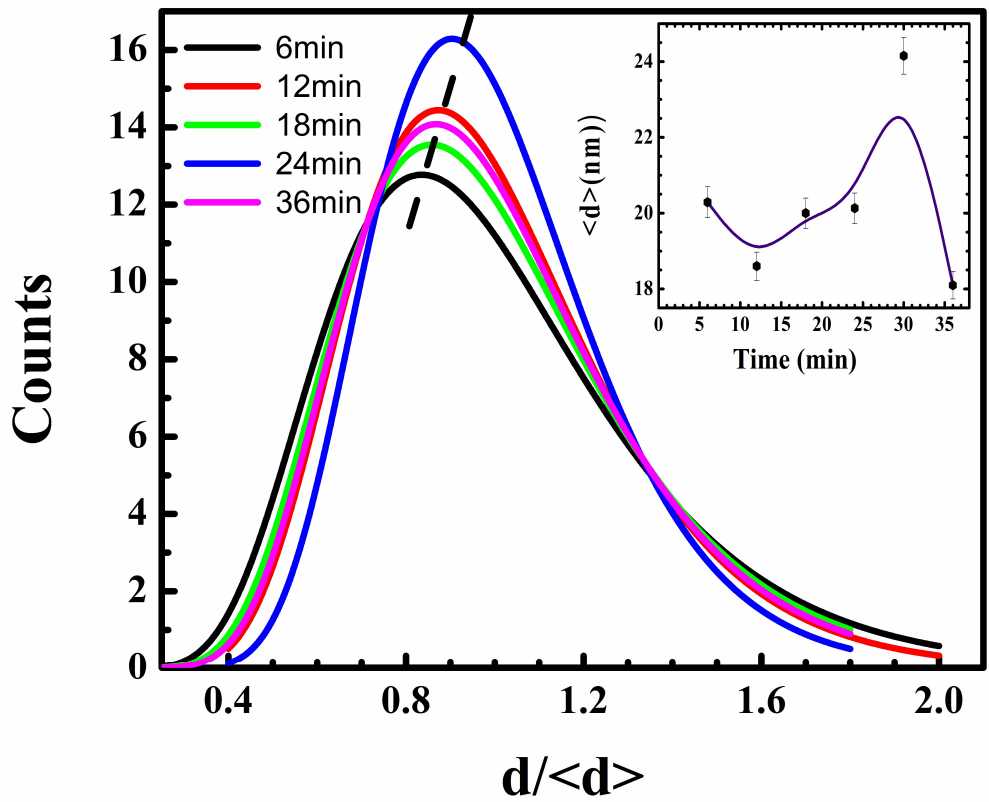

Figure5. Temporal evolution of the log-normal distribution observed from the TEM images. The inset reveals the temporal evolution of average particle size.

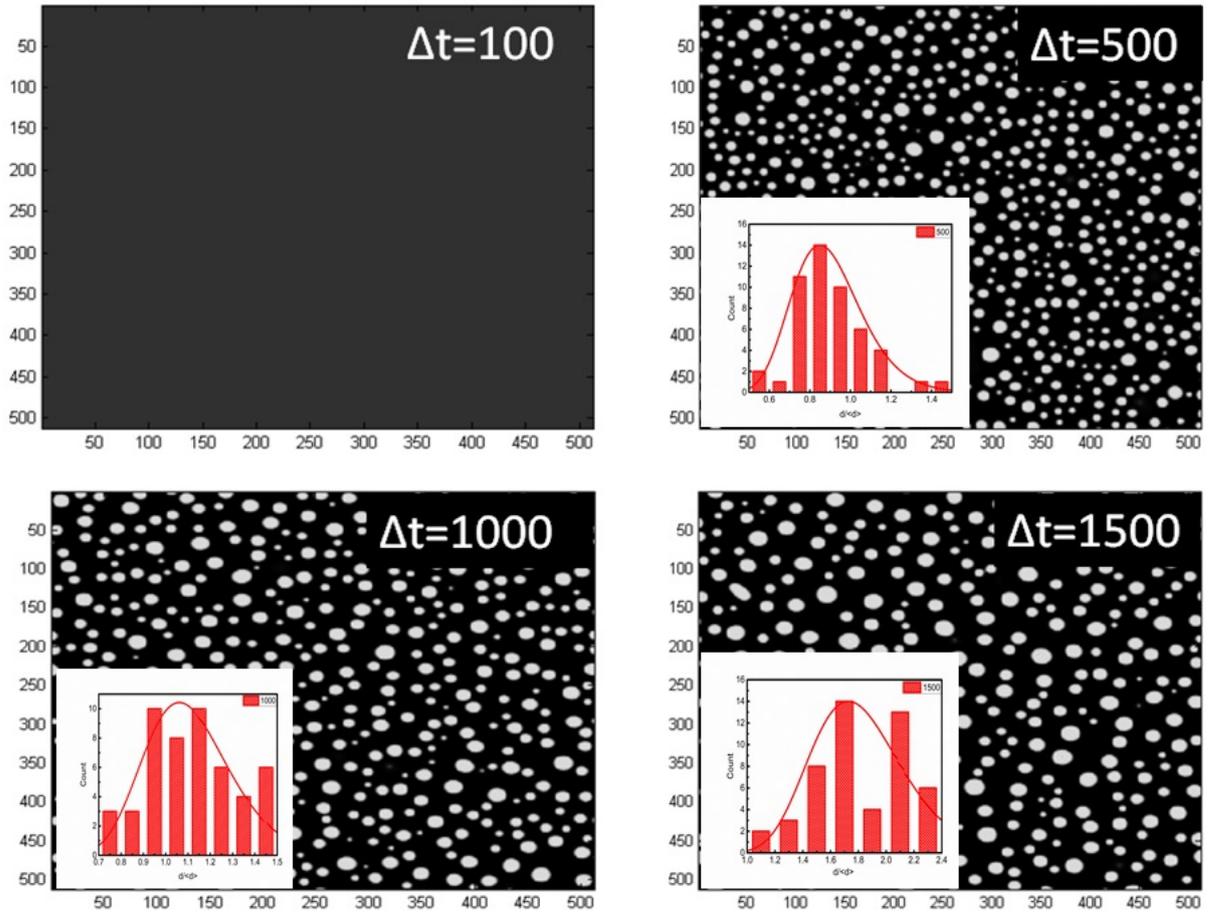

Figure 6. Simulated images and log-normal distribution (in inset) showing the temporal evolution of the nano-particles with different time steps having parameters (C = 0.25, A = 2, M = 0.5, κ = 0.5).

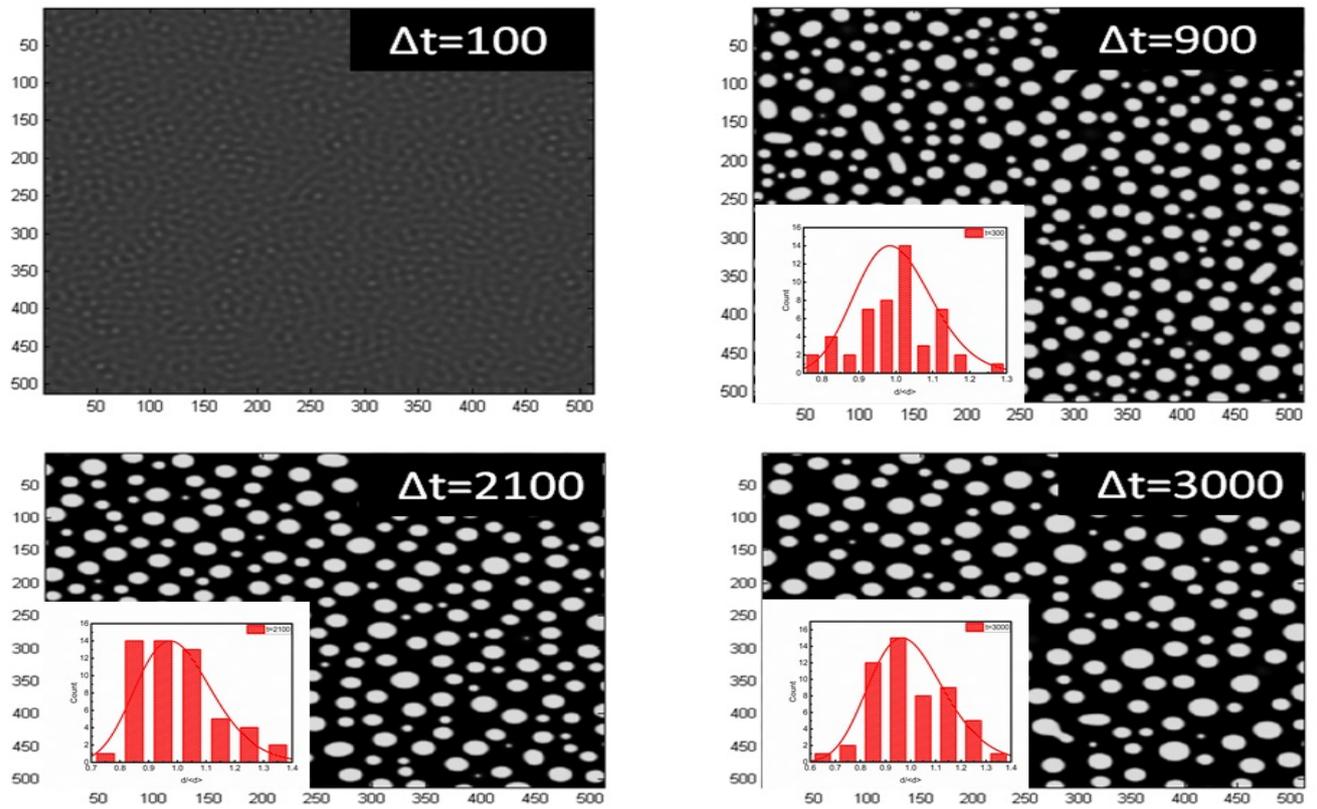

Figure7. Simulated images and log-normal distribution (in inset) showing the temporal evolution of the nano-particles with different time steps having parameters (C = 0.30, A = 1, M = 1, κ = 0.5).